# Stage-by-stage Wavelet Optimization Refinement Diffusion Model for Sparse-View CT Reconstruction

Kai Xu, Shiyu Lu, Bin Huang, Weiwen Wu, *Member, IEEE*, Qiegen Liu, *Senior Member, IEEE*

*Abstract*—Diffusion models have emerged as potential tools to tackle the challenge of sparse-view CT reconstruction, displaying superior performance compared to conventional methods. Nevertheless, these prevailing diffusion models predominantly focus on the sinogram or image domains, which can lead to instability during model training, potentially culminating in convergence towards local minimal solutions. The wavelet transform serves to disentangle image contents and features into distinct frequency-component bands at varying scales, adeptly capturing diverse directional structures. Employing the Wavelet transform as a guiding sparsity prior significantly enhances the robustness of diffusion models. In this study, we present an innovative approach named the Stage-by-stage Wavelet Optimization Refinement Diffusion (SWORD) model for sparse-view CT reconstruction. Specifically, we establish a unified mathematical model integrating low-frequency and high-frequency generative models, achieving the solution with optimization procedure. Furthermore, we perform the low-frequency and high-frequency generative models on wavelet's decomposed components rather than sinogram or image domains, ensuring the stability of model training. Our method rooted in established optimization theory, comprising three distinct stages, including low-frequency generation, high-frequency refinement and domain transform. Our experimental results demonstrate that the proposed method outperforms existing state-of-the-art methods both quantitatively and qualitatively.

*Index Terms*—Computed Tomography, Sparse-view, Image reconstruction, Diffusion model, Wavelet transform.[i]

## I. INTRODUCTION

Computed tomography (CT) plays a vital role in clinical diagnosis, industrial applications, and various other fields [1]. However, the cumulative radiation doses from repeated X-ray CT examinations raise concerns about potential health risks, including the development of cancer and other diseases [2]. As image quality is directly impacted by radiation dose [3], researchers have been actively exploring methods to enhance image quality in low-dose CT scanning. One representative approach for reducing radiation dose involves reducing the number of projections [4]. Sparse-view CT reconstruction is an attractive solution to address this issue. However, the presence of missing projections results in a notable degradation of image quality, resulting in streaking artifacts.

Sparse-view CT reconstruction is a challenging inverse problem, classical iterative reconstruction [5, 6] improves reconstruction quality with incomplete data over FBP algorithms. However, for highly sparse views, artifacts can occur on the reconstructed image. Compressed sensing (CS) [7, 8] provides a promising solution for incomplete projection reconstruction, utilizing sparsity priors such as total variation (TV) [9], wavelet frame [10], and dictionary learning [11]. Wavelet transform is a powerful signal analysis tool, known for its excellent time-frequency domain characteristics. Additionally, wavelet frame demonstrates a strong capability to sparsely estimate piece-wise smooth functions. Sahiner *et al.* [12] proposed a wavelet-based algorithm for restoring high-resolution wavelet images using approximate partial edge knowledge.

Deep learning-based reconstruction methods demonstrate great potential in terms of globalized artifacts suppression and computational cost reduction[13-16]. While supervised methods have shown good reconstruction results, their generality is one of many concerns. Surprisingly, diffusion models, renowned for their ability to perform exact sampling from complex data distributions [17]. This advantage extends successfully to sparse-view CT reconstruction. Guan *et al.* [18] introduced a fully score-based generative model in the sinogram domain for sparse-view CT reconstruction. Subsequently, Xia *et al.* [19] developed a patch-based denoising diffusion probabilistic model (DDPM). Furthermore, Xia *et al.* extended this approach to sub-volume-based 3D DDPM [20]. The applications of diffusion models focus on learning the score function in the sinogram or image domains. Clinical images typically contain abundant details and features at different scales, posing a challenge in capturing the diverse aspects. Additionally, the data distribution in medical imaging varies significantly across different sites and body positions, making it difficult to accurately model with a single generative model. These factors contribute to the instability observed in learned diffusion models within image or sinogram domains, ultimately compromising the quality of image reconstruction.

Wavelet transform is capable of decomposing an image into multi-frequency components, similar to other frequency decomposition methods. However, what sets it apart is its exceptional ability to preserve the spatial correspondence and directional properties between the transformed and original images [21]. As a typical sparsity prior, the wavelet transform is well-suited for extracting general features and structures. The first property of the wavelet transform facilitates the separation of image contents and features into different frequency-component bands at various scales. This allows for a more effective extraction of diverse features. Additionally,

This work was supported in part by National Natural Science Foundation of China under 62122033, and Key Research and Development Program of Jiangxi Province under 20212BBE53001. (K. Xu and S. Lu are co-first authors.) (Corresponding authors: Weiwen Wu and Qiegen Liu.)
This work did not involve human subjects or animals in its research.
K. Xu, S. Lu, and B. Huang, are with School of Mathematics and Computer Sciences, Nanchang University, Nanchang 330031, China ({xukai, lushiyu, huangbin}@email.ncu.edu.cn).
W. Wu is with the School of Biomedical Engineering, Sun Yat-Sen University, Shenzhen, Guangdong, China (wuweiw7@mail.sysu.edu.cn).
Q. Liu are with School of Information Engineering, Nanchang University, Nanchang 330031, China. ({liuqiegen}@ncu.edu.cn).

the second property enhances the capability of capturing different directional structures, further enriching the information extracted from the image. Furthermore, by incorporating the wavelet transform as a prior to constrain the generative result, it efficiently improves the robustness of the diffusion models. In fact, deep learning-based models that utilize directional wavelets have demonstrated effectiveness in preserving essential information [12]. For instances, Bae *et al.* [22] introduced a novel feature space deep residual wavelet learning algorithm for restoring image details, while Liu *et al.* [23] designed a multi-level wavelet-CNN that effectively recovers detailed textures and sharp structures. These findings further highlight the advantages of the wavelet transform.

Inspired by the aforementioned facts, integrating the wavelet transform into diffusion models proves to be advantageous in stabilizing the model's training process. By dividing the image into high-frequency and low-frequency components, the score function of the low-frequency component focuses on learning the main structures and features, while the score function from the high-frequency component concentrates on small details and features. Hence, it becomes crucial to handle them separately to achieve better stability and accuracy during training.

Therefore, we propose a novel Stage-by-stage Wavelet Optimization Refinement Diffusion (SWORD) model for sparse-view CT reconstruction. The training of the model comprises three consecutive stages: low-frequency generation, high-frequency refinement and domain transform. In the first stage of low-frequency generation, the wavelet-based full-frequency diffusion model (WFDM) serves as the initial generator, taking advantage of the sparsity of wavelet transform to explore main features and profile information comprehensively. The second stage of the low-frequency components refinement employs the wavelet-based high-frequency diffusion model (WHDM) to refine high-frequency structures and features with focusing solely on high-frequency information. As a result, WHDM acquires richer details, delivering sufficient texture and clear details. In the last stage, we reconstruct the final images using domain transforms, including Wavelet inverse transform and filtered back-projection. The key contributions of this work are summarized as follows:

● We present a pioneering wavelet-domain diffusion model, operating within the wavelet domain instead of the original data or image domains. This innovative approach substantially improves the stability of the diffusion model throughout the training process. By leveraging the wavelet transform, our model can effectively capture and represent features and structures within sinogram.

● By modeling the sinogram using both high-frequency and low-frequency components with a unified mathematical model, we effectively separate the complex data distribution into two independent simplified distributions. Consequently, two score functions are incorporated into sinogram reconstruction to characterize these two distinct data distributions by optimizing the established mathematical model. This innovative method allows us to address the challenges posed by data complexity.

● We also develop a two-stage wavelet-based diffusion strategy. In the first stage, the focus is on learning the low-frequency components that encompass the main structures and features of the sinogram. In the second stage, the emphasis is on constructing a generative model for the high-frequency components to capture intricate details and structures. These two-stage approach enables the model to learn and leverage both global and local information effectively, contributing to enhanced reconstruction quality and accuracy.

● We rigorously validate and assess our proposed model on two large-scale sparse-view CT datasets. The experimental results showcase its remarkable reconstruction performance, substantiated by both quantitative and qualitative evaluations. The model's proficiency in producing high-quality reconstructions reaffirms its potential as a robust and effective solution for sparse-view CT reconstruction.

The remaining sections of the manuscript are structured as follows. In Section II, we provide a brief overview of relevant works. Section III establishes a unified theory to present a comprehensive explanation of our proposed method. Section IV showcases the experimental results with implementation specifications. Finally, we draw the conclusions.

## II. PRELIMINARY

### A. CT Imaging Model

Image reconstruction from incomplete data poses an inherently ill-posed inverse problem, presenting substantial challenges in achieving accurate solutions. Sparse-view data reconstruction is a highly challenging topic in the field of CT imaging. The linear inverse problem of conventional CT imaging can be described as follows:

$$\boldsymbol{x} = \boldsymbol{AI}, \quad (1)$$

where $\boldsymbol{x}$ represents the acquired full projection data and $\boldsymbol{I}$ represents the reconstructed image. $\boldsymbol{A}$ is the imaging system matrix, it is non-invertible in sparse-view CT reconstruction. Hence, this inverse problem is highly ill-posed. To tackle this challenge, the regularization-based models have been extensively adopted, as explained below:

$$\min_{\boldsymbol{I}} \|\boldsymbol{AI} - \boldsymbol{x}\|_2^2 + \lambda R(\boldsymbol{I}), \quad (2)$$

where the first term is the data consistency, and $R(\boldsymbol{I})$ is the regularization term that incorporates the prior knowledge of the reconstructed image $\boldsymbol{I}$.

### B. Diffusion Models

Diffusion models have demonstrated remarkable performance in various generative modeling tasks. They can be primarily categorized into two types: denoising diffusion probabilistic models (DDPMs) [24, 25] and stochastic differential equations (SDEs) [26].

**DDPMs**: DDPMs can be regarded as a specialized form of variational autoencoders [27]. For instance, Sohl *et al.* [24] introduced a novel algorithm for modeling probability distributions by estimating the reverse of a Markov diffusion chain, which maps data to a noise distribution. Building upon considerations from nonequilibrium, Ho *et al.* [25] presented the DDPM by estimating the image noise at each reverse process step. Song *et al.* [28] proposed the denoising diffusion implicit model (DDIM) by replacing the Markov forward process used in [24] with a non-Markovian one. Nachmani *et al.* [29] suggested substituting the Gaussian noise distributions of the diffusion process with a mixture of two Gaussians or the Gamma distribution for faster convergence.

**SDEs**: Song *et al.* [26] initially introduced the stochastic differential equation (SDE), comprising the forward diffusion process and its corresponding reverse-time SDE. Wang *et al.* [30] presented a novel deep generative model based on Schrödinger bridge. To address the limitations of high-dimensional score-based models caused by Gaussian noise distribution, Deasy *et al.* [31] extended the denoising score matching method to incorporate heavier-tailed distributions such as normal noising distribution. By encoding the image in the latent space, Kim *et al.* [32] proposed a non-linear diffusion process employing a trainable normalizing flow model.

*C. Wavelet-based Reconstruction Approaches*

The wavelet tight frame has gained widespread recognition for its capacity to sparsely approximate piecewise-smooth functions [33, 34], making it a widely adopted system. In the context of sparse view CT reconstruction, Qu *et al.* [35] introduced a hybrid reconstruction model that effectively addresses the problem by incorporating TV regularization with the wavelet frame. Similarly, Mehranian *et al.* [36] proposed a projection completion based metal artifact reduction algorithm, formulating the completion of missing projections as a regularized inverse problem in the wavelet domain. Unfortunately, above methods simply use hand-designed filters or objective functions to reconstruct sparse-view CT image, which dissatisfies the real image priors and often leads to poor results.

The integration of wavelet transform into deep learning-based systems has proven successful, benefiting from its ability to leverage spatial and frequency information during training. For instance, Lee *et al.* [37] presented a novel approach to sparse-view CT reconstruction, utilizing a multi-level wavelet convolutional neural network to enhance performance and accuracy. Furthermore, Bai *et al.* [38] introduced a hybrid domain method that combines wavelet, spatial, and Radon domains priors for sparse-view CT reconstruction. Although deep learning-based methods present impressive performance, the results are still unsatisfactory.

## III. METHOD

This section provides a detailed description of the proposed Stage-by-stage Wavelet Optimization Refinement Diffusion (SWORD) model. Firstly, we elaborate on the motivation behind SWORD. Next, we introduce the core of wavelet-based forward diffusion process within two stages, which enables more efficient sampling. Additionally, we develop a unified mathematical model to perform optimization refinement reconstruction.

*A. Motivation*

The sparsity property of the wavelet framework enables the extraction of meaningful features during training, leading to significant improvements in preserving key information while requiring fewer coefficients to represent the data. Furthermore, the wavelet transform offers topologically simpler manifold structures while retaining directional edge information, facilitating the model's ability to learn sufficient prior information in the wavelet domain. Leveraging this advantage, introducing noise in the wavelet domain enables more stable and accurate reconstruction processes.

In typical measurements, both high-frequency and low-frequency components are present. The high-frequency component captures intricate textures and detailed information, while the low-frequency component represents global structures. During denoising, preserving high-frequency information is crucial. By dividing the image into high and low-frequency components, the model can focus on learning structures and features separately. Training them jointly can lead to instability. Therefore, handling them separately ensures better stability and accuracy during training.

The wavelet transform provides a means to separate data into four sub-bands, including the high-frequency and low-frequency components. By extracting the high-frequency component individually, as shown in Fig. 1, we can preserve the edge information and texture details. To achieve this, the low-frequency sub-band is replaced by an all-zero matrix. Subsequent operations, such as the inverse wavelet transform ($W^T$) and the FBP, clearly reveal the edge information and texture details in both the sinogram and image domains.

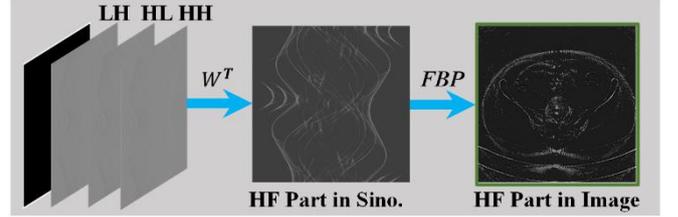

Fig. 1. Reconstruction result of high-frequency data, where HF represents high-frequency.

In this study, we propose a stage-by-stage wavelet-based optimization refinement diffusion strategy aimed at enhancing image quality and accuracy while providing a comprehensive perspective for the diffusion model. The main process of the generation includes two stages, which is depicted in Fig. 2. The first stage, known as the wavelet-based full-frequency diffusion model (WFDM), leverages the sparsity of Wavelet transform to extract main features and profile information from the data. In the second stage, denoted as the wavelet-based high-frequency diffusion model (WHDM), the high-frequency components obtained from the first stage undergo further refinement. This two-stage approach utilizes two trained score functions to model these distinct data distributions, resulting in precise and robust reconstruction outcomes with rich texture and clear details. This approach ensures improved sinogram fidelity and the preservation of essential information.

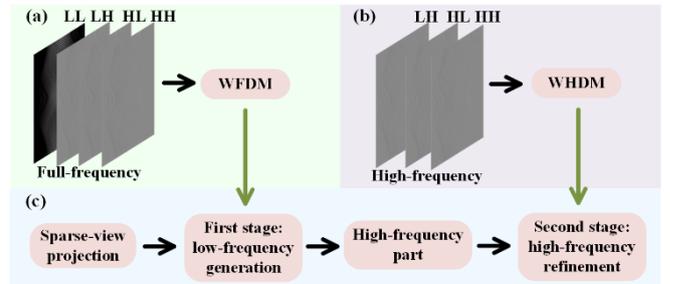

Fig. 2. The main process of the two-stage strategy. (a) The simple training process for WFDM. (b) The simple training process for WHDM. (c) The simple test process for SWORD.

*B. SWORD Model*

*1) Problem Formulation:* The transition from full projection data to sparse-view data can be interpreted as a linear measurement process. Fig. 3 provides an intuitive visualization of this linear measurement process. By applying the Radon transform, the original image is transformed into the projection data. Additionally, $diag(\Lambda)$ represents the subsampling mask for the projection data, and $P(\Lambda)$ subsamples the sinogram to sparse-view data with subsampling mask. Consequently, the sparse-view CT reconstruction problem can be formulated as:

$$\boldsymbol{y} = P(\Lambda)\boldsymbol{AI} = P(\Lambda)\boldsymbol{x}, \qquad (3)$$

where $\boldsymbol{y}$ is the sparse-view CT projection data.

When dealing with sparse-view CT data, directly applying the FBP algorithm to reconstruct an image often results in strip artifacts. To achieve higher quality reconstructed images, it is necessary to derive full-view projection data from sparse-view projection data, which poses an underdetermined inverse problem. To address this challenge, the priors can be incorporated into a regularized objective function with minimizing the following problem:

$$x^* = \arg\min_{x} \|P(\Lambda)x - y\|_2^2 + \gamma R(x), \quad (4)$$

The objective function consists of two terms. The first term represents the data fidelity, ensuring that the full-view projection data aligns with the measurements obtained through the subsampling mask. The second term, is the regularization term, where $\gamma$ helps maintain an appropriate balance between data fidelity and regularization.

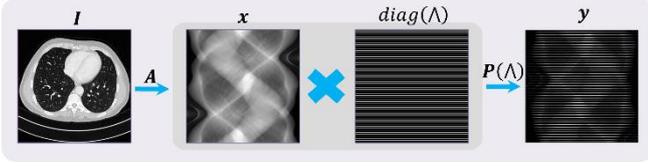

**Fig. 3.** Linear measurement process for sparse-view CT.

*2) Wavelet-based Forward Diffusion:* Acknowledging the advantages of incorporating wavelet transform into diffusion models to enhance training stability, we present an innovative strategy wherein the diffusion process functions on the outcomes of wavelet transformation, departing from the utilization of the original sinogram. By decomposing the projection data into four distinctive sub-bands through wavelet transformation, the data undergoes an effectively sparse processing technique.

Our solution is grounded in discrete wavelet transform (i.e., DWT), a technique that dissects the sinogram into four distinct sub-bands: the low-frequency (LL) and the high-frequency (LH, HL, HH) components, as illustrated in Fig. 4. The LL component encapsulates the principal features and structures, while the HL, LH, and HH components encapsulate detailed high-frequency elements in vertical, horizontal, and diagonal orientations respectively. This characteristic DWT operation supplies the model with more comprehensive statistical insights, facilitating the acquisition of more advantageous priors and an augmentation of its capabilities. Consequently, our approach yields notable enhancements in model performance without compromising output quality to a significant degree. The construction of the four sub-bands image, achieved by applying DWT to the sinogram data of training sample, can be expressed using the subsequent equation:

$$W(x) = a_{LL}(x) + d_{LH}(x) + d_{HL}(x) + d_{HH}(x), \quad (5)$$

where $W$ is denoted as wavelet transform. $a_{LL}(x)$ and $\{d_{LH}(x), d_{HL}(x), d_{HH}(x)\}$ are denoted as the low-frequency (LL) component and the high-frequency (LH, HL, HH) elements respectively. To streamline and expedite our explanation, we introduce the shorthand notation, $X_1 = \{a_{LL}(x), d_{LH}(x), d_{HL}(x), d_{HH}(x)\}$.

While wavelet-based reconstruction approaches yield commendable results, they tend to fall short in preserving fine intricacies. Recognizing that high-frequency information is synonymous with intricate details, we prioritize the accurate capture of these high-frequency components by devising a distinct diffusion model solely dedicated to handling high-frequency data. Within this model, we focus on the selection of three specific high-frequency sub-bands, namely $X_2 = \{d_{LH}(x), d_{HL}(x), d_{HH}(x)\}$, aiming to glean more comprehensive prior insights concerning image intricacies.

Hence, the training of SWORD model consists of two stages: WFDM and WHDM. The WHDM is primed through training with all four complete-frequency components as an initial generation, while the WFDM hones its prowess by being trained with the three high-frequency components to refine the image edges and features within high-frequency band. The WFDM ingeniously harnesses the wealth of high-frequency information to amplify intricate details and accentuate features within the initially generated images. Through the orchestrated collaboration of WHDM and WFDM in a sequential orchestration, the SWORD model adeptly amalgamates an extended spectrum of prior knowledge, thereby instigating noteworthy advancements in the caliber of the reconstructed images.

Regarding as the diffusion model, we empirically utilize the score-based model, specifically the Stochastic Differential Equation (SDE) in this study, to train the diffusion model within the two stages. The score-based model involves both the forward SDE and the reverse SDE. Fig. 4 provides a detailed depiction of these processes in the wavelet domain. Throughout the training procedure, the forward SDE transforms the intricate data distribution into a known prior distribution by gradually introducing noise. Consequently, the SDE method could handle a continuous process.

Consider a continuous diffusion process $\{x(t)\}_{t=0}^T$ where $x(t) \in \mathbb{R}^n$, with time index $t \in [0, T]$ marking the progression and $n$ representing the image dimension. In this context, the forward SDE process can be expressed as:

$$dx = f(x, t)dt + g(t)dw, \quad (6)$$

Here, $f(x, t) \in \mathbb{R}^n$ signifies the drift coefficient, and $g(t)$ represents the diffusion coefficient. The term $w \in \mathbb{R}^n$ introduces Brownian motion.

To be specific, we embrace the variance exploding (VE) SDE configuration ($f = 0$, $g = \sqrt{d[\sigma^2(t)]/dt}$), a choice that imparts heightened generative prowess. Upon incorporating $X_1$ and $X_2$ into Eq. (6), the equation metamorphoses into the following forms:

$$dX_1 = \sqrt{d[\sigma^2(t)]/dt}\, dw, \quad (7)$$
$$dX_2 = \sqrt{d[\sigma^2(t)]/dt}\, dw, \quad (8)$$

where the $\sigma(t) > 0$ denotes a monotonically increasing function, thoughtfully chosen to manifest as a geometric series.

Suppose $X_1(0)$ follows the distribution $p_0$, representing the initial data distribution of wavelet transform components, while $X_1(T)$ follows $p_T$, symbolizing the acquired prior distribution. Throughout the training phase, the efficacy of the neural network is finely tuned through variations in the parameter $\theta_1^*$. This can be perceived as the core objective function within the context of the score-based SDE:

$$\theta_1^* = \arg\min_{\theta_1} \mathbb{E}_t\{\lambda(t)\mathbb{E}_{X_1(0)}\mathbb{E}_{X_1(t)|X_1(0)}[\|s_{\theta_1}(X_1(t), t) - \nabla_{X_1(t)} \log p_t(X_1(t)|X_1(0))\|_2^2]\}, \quad (9)$$

Here, $\lambda(t)$ represents a positively weighted function, and $t$ is uniformly sampled from the interval $[0, T]$. The term $p_t(X_1(t)|X_1(0))$ signifies the Gaussian perturbation kernel centered around $X_1(0)$. A neural network $s_{\theta_1}$ is systematically trained to approximate the score, allowing it to capture the essence of $\nabla_{X_1} \log p_t(X_1)$, i.e., $s_{\theta_1} = \nabla_{X_1} \log p_t(X_1)$. Upon achieving full model proficiency, the value of $\nabla_{X_1} \log p_t(X_1)$ can be effectively determined across all time points $t$ through the solution of $s_{\theta_1}(X_1(t), t)$. This equips the model with an ample reservoir of prior knowledge, thereby enabling it to skillfully undertake the process of image reconstruction. In fact, the training of $X_2$ can be following the same procedures to obtain $s_{\theta_2}$.

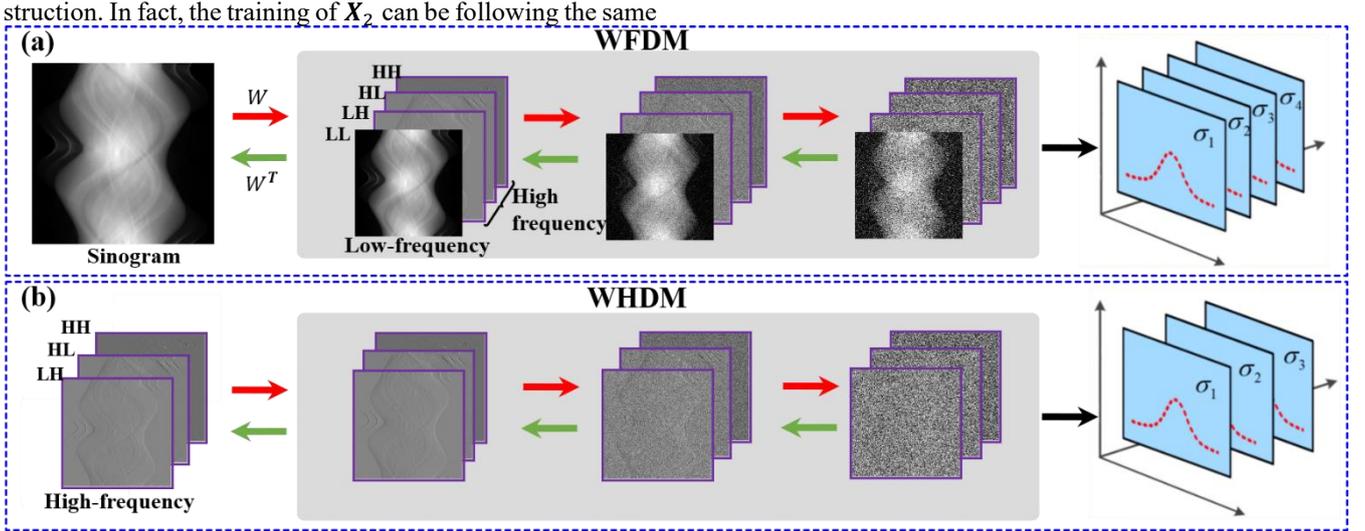

**Fig. 4.** Visual Representation of forward diffusion procedures of the proposed SWORD. The projection data undergo a transformation into distinct sub-bands, encompassing a singular low-frequency sub-band (LL) alongside three high-frequency sub-bands (HH, HL, and LH). The Forward Wavelet-enhanced Diffusion Model (WFDM) is trained using all four sub-bands, while the High-frequency Wavelet Diffusion Model (WHDM) is exclusively trained with the three high-frequency sub-bands. From top to bottom: (a) The elaborate training process for WFDM. (b) The elaborate training process for WHDM.

## C. Optimization Refinement Reconstruction

The comprehensive process of reconstruction can be efficiently executed by orchestrating a physical-information inspired conundrum. This entails the harmonious integration of regularization priors targeting both low-frequency components to preserve inherent structure, and high-frequency details to uphold intricate nuances. The overarching optimization objective is elegantly framed as the minimization of the subsequent expression:

$$\{X_1^*, X_2^*\} = \underset{\{X_1, X_2\}}{argmin}[\|X_1 - WP(\Lambda)y\|_2^2 + \lambda_1\|X_2 - E(X_1)\|_2^2 + \lambda_2 R_1(X_1) + \lambda_3 R_2(X_2)], \quad (10)$$

where the hyperparameter $\lambda_1$ plays a pivotal role in striking an equilibrium between the influences of high-frequency components. It can be likened to the data fidelity governing the comportment of high-frequency compounds. Meanwhile, $\lambda_2$ and $\lambda_3$ are regularization factors, diligently orchestrating a delicate balance between the imperatives of data consistency elements and the guiding principles of regularization priors. The $E(\cdot)$ represents the extractor of high-frequency component within wavelet domain, where $E^T E(X_1) = X_1$.

Within the construct of Eq. (10), it becomes apparent that the optimization function harbors a symphony of two data fidelities and a duo of regularization priors. The initial data consistency stems from the unaltered sinogram, while the second arises from the alignment of high-frequency data consistencies. This optimization endeavor unfurls in a choreography of two consecutive stages, each contributes distinctively to the refinement process:

$$\begin{cases} X_1^* = \underset{X_1}{argmin}\left[\|X_1 - WP(\Lambda)y\|_2^2 + \left\|\lambda_1 X_2^{i+1} - E(X_1)\right\|_2^2 + \lambda_2 R_1(X_1)\right] \\ X_2^* = \underset{X_2}{argmin}\left[\|X_2 - E(X_1^i)\|_2^2 + \lambda_3 R_2(X_2)\right] \end{cases}, \quad (11)$$

where $i$ represents the current iteration step, ranging from $T - 1$ to $0$. The evolution of reconstruction results occurs in a dual rhythm, transitioning from its initial phase to a refined stage. This sequential progression is orchestrated harmoniously by WFDM and WHDM. WHDM draws wisdom from four encompassing frequency sub-bands, fulfilling its role, while WFDM refines its efficacy by skillfully adjusting three polished high-frequency sub-bands. Actually, Eq. (11) can be further optimized with updating the $X_1$ and $X_2$ with an iteration strategy.

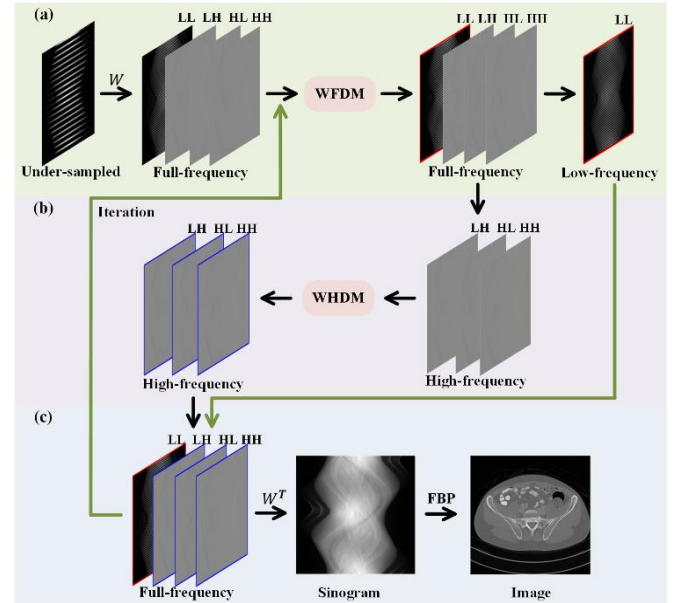

**Fig. 5.** The reconstruction phase of SWORD in sinogram domain for sparse-view CT. The generation process can be mainly divided into three stages: $X_1$ and $X_2$ and backprojection stages. WT denotes the wavelet transform. From top to bottom: (a) Low-frequency generation stage, (b) High-frequency refinement stage, (c) Domain transform stage.

*Low-frequency Generation:* Regarding as the sub-problem, it can be treated as low-frequency generation. $X_1$ can be further decomposed into two sub-problem as following:

$$X_1^{i+1/2} = \underset{X_1}{argmin}\|X_1 - WP(\Lambda)y\|_2^2 + \lambda_1\|X_2^{i+1} - E(X_1)\|_2^2, \quad (12a)$$

$$X_1^i = \underset{X_1}{argmin}\left[\|X_1 - X_1^{i+1/2}\|_2^2 + \lambda_2 R_1(X_1)\right], \quad (12b)$$

In the Eq. (12a), it can be considered as data consistency from sinogram and can be solved with derivative decent

method with following expression:
$$X_1^{i+1/2} = X_1^{i+1} - \eta(X_1^{i+1} - WP(\Lambda)y + \lambda_1 E^T(E(X_1^{i+1}) - X_2^{i+1})), \quad (13)$$

where $\eta > 0$ is step length. From Eq. (13), we find that the contribution of $\lambda_1 E^T X_2^{i+1}$ is small compared with $WP(\Lambda)y$, since the former focus on the high-frequency components and the later only concentrates on the full components. Here, we can further simplify the Eq. (13) with
$$X_1^{i+1/2} = X_1^{i+1} - \eta(X_1^{i+1} - WP(\Lambda)y), \quad (14)$$

In respect to the Eq. (12b), it is one of SDE inverse solving problem incorporating the diffusion prior. Upon the training of the WFDM model, the acquisition of $\nabla_{X_1} \log p_t(X_1)$ enables the estimation of the prior distribution of projection data in Wavelet domain. The WFDM is then leveraged to generate low-frequency components, thereby ensuring the establishment of a fundamental reconstruction baseline. The subsequent Predictor and Corrector (PC) operation collectively encapsulate the core of generation process.

The execution involves the Predictor-Corrector (PC) solver, which comprises two distinct components: the Predictor and the Corrector. The Predictor is responsible for sampling from the instantaneous data distribution through a numerical SDE. In contrast, the Corrector aims to refine the Predictor's outcomes using the Monte Carlo Markov Chain (MCMC) technique. The discretization of the numerical SDE solver can be expressed as follows:
$$X_1^i \leftarrow X_1^{i+1/2} + (\sigma_{i+1}^2 - \sigma_i^2) s_{\theta_1}(X_1^{i+1/2}, \sigma_{i+1}) + \sqrt{\sigma_{i+1}^2 - \sigma_i^2} z, \quad i = T-1, \ldots, 0 \quad (15)$$

where $z$ follows a standard normal distribution $z \sim N(0,1)$, and $X_1(0)$ originates from distribution $p_0$ with $\sigma_0 = 0$ for brevity.

*High-frequency Refinement:* This stage can be called as high-frequency refinement stage. Regarding for solving Eq. (12b), the high-frequency sub-band of $X_1$ is treated as $X_2$, i.e., the $a_{LL}(x)$ within $X_1$ is discarded. We have achieved the data gradient $\nabla_{X_2} \log p_t(X_2)$ with training the WHDM model, enabling the estimation of the underlying distribution high-frequency features in the projection data. This WHDM model serves as a means to generate high-frequency features, which is benefit for details and feature recovery. While both WFDM and WHDM operate within the wavelet domain, their primary distinction lies in WHDM's emphasis on high-frequency components within the sub-bands. Given that high-frequency information corresponds to the image's finer intricacies, WHDM excels at creating intricate and textured details.

Analogous to the $X_1$ stage, the updating of $X_2$ utilizes the WHDM training model as a regularization, where the PC operation is similar to Eq. (15). Following its passage through the WHDM model, the $X_2$ undergoes an update. The updated $X_2^{i-1}$ is then amalgamated with the previously reserved $a_{LL}(x^{i-1})$ within $X_1$ to generate the $X_1$.

Through a continuous iterative progression of stage-to-stage generation, the intricate details and edges of images can be comprehensively reconstructed, all the while upholding the foundational essence of reconstruction. By sequentially integrating WFDM and WHDM, the SWORD methodology capitalizes on high-frequency information to delve deeper into the reservoir of prior knowledge concerning image details, thereby achieving a heightened degree of accuracy in reconstruction.

*Domain Transform Stage:* When the iteration is over, the reconstruction result can be acquired by combining Wavelet inverse transform operation and FBP algorithm from the combination of $X_1$ and $X_2$ with following expression
$$\tilde{I} = FBP(W^T(\{X_1 - E(X_1), X_2\}, X_2)) \quad (16)$$
where $W^T$ is the inverse wavelet transform, $X_1 - E(X_1)$ signifies the low-frequency component. $\tilde{I}$ stands as the final reconstruction results produced by our SWORD method. To summarize, the flowchart of the iterative reconstruction phase for SWORD is depicted in Fig. 5. Specifically, two wavelet-based models are trained to capture the prior distribution. Then, during the inference stage, the numerical SDE solver and MCDM steps are iteratively updated to obtain the complete projection data for a full-view projection. The introduction of DC ensures the convergence of SWORD. Additionally, Algorithm 1 provides a detailed description of the training process and the inference stage.

| **Algorithm 1: SWORD for iterative generation** |
|---|
| **Training Process** |
| 1: Generate projection with $x = AI$; |
| 2: Wavelet transform (WT): $W(x) = a_{LL}(x) + d_{LH}(x) + d_{HL}(x) + d_{HH}(x)$; |
| 3: Training datasets construction: $X_1 = \{a_{LL}(x), d_{LH}(x), d_{HL}(x), d_{HH}(x)\}$; $X_2 = \{d_{LH}(x), d_{HL}(x), d_{HH}(x)\}$; |
| 4: Training with Eq. (9); |
| 5: Output: $s_{\theta_1}(X_1, t)$ and $s_{\theta_2}(X_2, t)$. |
| **Inference Process** |
| **Setting:** $s_{\theta_1}, s_{\theta_2}, T, \sigma, \varepsilon$ |
| 1: $X_1^T \sim N(0, \sigma_{max}^2), X_2^T \sim N(0, \sigma_{max}^2)$ |
| 2: For $i = T - 1$ to $0$ **do** |
| 3:   Update $X_1^i \leftarrow Predictor(X_1^{i+1}, \sigma_i, \sigma_{i+1})$; |
| 4:   Update $X_1^i$ by data consistency with Eq. (12a); |
| 5:   Update $X_1^i \leftarrow Corrector(X_1^i, \sigma_i, \varepsilon_i)$; |
| 6:   Update $X_1^i$ by data consistency with Eq. (12a); |
| 7:   Repeat steps 3-6 by replacing $X_1$ with $X_2$; |
| 8: End for |
| 9: Achieving the reconstructed image by Eq. (16); |
| 10: Return $\tilde{I}$. |

## IV. EXPERIMENTS

### A. Data Specification

**AAPM Challenge Data:** The dataset used for evaluation is simulated from human abdomen images, generously provided by Mayo Clinic for the AAPM Low Dose CT Grand Challenge [39]. The dataset comprises high-dose and low-dose CT scans from 10 patients, with 9 used for training and 1 for evaluation. Here, we employed 5388 slices with a thickness of 1 mm, each covering a total of pixels. Among these, 4839 images are used for training, and 549 images are used for testing. Reference images are generated from 720 projection views using the FBP algorithm.

For sparse-view CT reconstruction, we extract 60, 90, 120, and 180 views data from the full-view projection. For fan-beam CT reconstruction, we utilize Siddon's ray-driven algorithm [40, 41] to generate the projection data. The distance from the rotation center to the source and detector is set to 40 cm and 40 cm, respectively. The detector width is 41.3 cm, with 720 detector elements, and the total 720 projection views are evenly distributed.

**CIRS Phantom Data:** We acquire a high-quality dataset of

CT volumes with dimensions of 512×512×100 voxels and a voxel size of 0.78×0.78×0.625 mm³. This dataset is obtained using an anthropomorphic CIRS phantom on a GE Discovery HD750 CT system, with the tube current set to 600 mAs. The CT system's source-to-axial distance is 573 mm, and the source-to-detector distance is 1010 mm. For sparse-view CT reconstruction, we extract 60, 90, 120, and 180 views data from the original full-view projection.

*B. Model Training and Parameter Selection*

In our experiments, we train the model using the Adam algorithm with a specific learning rate $10^{-3}$, and we initialize the weights using the Kaiming initialization method. The implementation language is programmed in Python, employing the Operator Discretization Library (ODL) [42] and PyTorch. The computations are performed on a workstation equipped with a GPU (NVIDIA GTX 1080Ti-11GB).

During the reconstruction stage, we set the number of iterations to 1450. For each execution of the prediction and correction process, we utilize the annealing Langevin. For those interested in exploring our method, the source code is publicly accessible at: https://github.com/yqx7150/SWORD. To quantitatively evaluate the performance, three standard metrics including peak-signal-to-noise ratio (PSNR), structural similarity index (SSIM), and mean squared error (MSE) are employed.

*C. Reconstruction Results*

In this section, we conduct experiments to address the sparse-view CT reconstruction task. To evaluate the effectiveness of our proposed algorithm, we compare its performance with other methods, including FBP [1], U-Net [43], FBPConvNet [44], Patch-based DDPM [45], and GMSD [18]. Here, we employ 60, 90, 120, and 180 views for the sparse-view CT reconstruction. FBP is a classic image reconstruction technique for sparse-view CT. U-Net is a widely-used neural network architecture for various image reconstruction tasks. FBPConvNet combines FBP and deep learning techniques to enhance CT reconstruction. Patch-based DDPM is based on diffusion model for low-dose CT reconstruction. GMSD is a state-of-the-art generative model known for generating high-quality images. Notably, U-Net and FBPConvNet belong to supervised methods, while Patch-based DDPM and GMSD operates in an unsupervised manner.

We perform these experiments on two different datasets: AAPM Challenge Data and CIRS Phantom Data. The reconstruction results of the baseline methods and our proposed approach are presented in Tables I-II, and visualized in Figs 6-9. The best PSNR, SSIM, and MSE values of the reconstructed images with different projection views are highlighted in bold in the tables. For clearer examination, we magnify the detailed region-of-interest (ROI).

*AAPM Reconstruction Results:* For sparse-view CT reconstruction using 60, 90, 120, and 180 projection views, we present the average PSNR, SSIM and RMSE values of the reconstructed results from the AAPM Challenge Dataset in Table I. The table reveals that our proposed method, SWORD, achieves substantially the highest PSNR, SSIM, and MSE values compared to other reconstruction methods. Notably, SWORD shows a remarkable improvement over the traditional FBP method. When compared to both supervised and unsupervised methods, SWORD consistently outperforms them, demonstrating its exceptional generalization and high performance as an unsupervised approach.

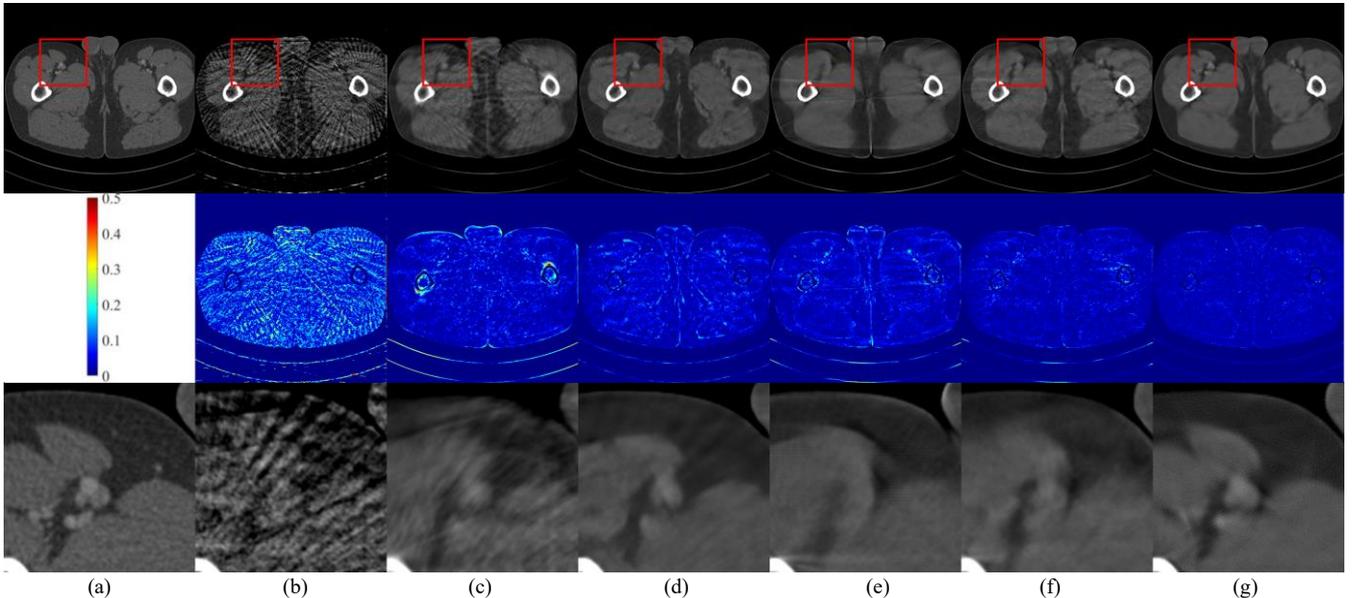

**Fig. 6.** Reconstruction images from 60 views using different methods. (a) The reference image versus the images reconstructed by (b) FBP, (c) U-Net, (d) FBPConvNet, (e) Patch-based DDPM, (f) GMSD, and (g) SWORD. The display windows are [-250,600]. The second row is residuals between the reference images and reconstruction images.

TABLE I RECONSTRUCTION PSNR/SSIM/MSE($10^{-3}$) OF AAPM CHALLENGE DATA USING DIFFERENT METHODS AT 60, 90, 120, AND 180 VIEWS.

| Views | FBP | U-Net | FBPConvNet | Patch-based DDPM | GMSD | SWORD |
|---|---|---|---|---|---|---|
| 60 | 23.18/0.5950/4.88 | 28.83/0.9365/1.56 | 35.63/0.9659/0.28 | 32.04/0.9336/0.68 | 34.31/0.9580/0.41 | **38.49/0.9778/0.15** |
| 90 | 26.20/0.7013/2.45 | 30.09/0.9472/1.17 | 37.11/0.9758/0.25 | 35.15/0.9634/0.35 | 37.25/0.9739/0.20 | **41.27/0.9862/0.08** |
| 120 | 28.30/0.7865/1.52 | 35.58/0.9765/0.34 | 39.45/0.9828/0.15 | 37.90/0.9759/0.17 | 39.41/0.9812/0.12 | **42.49/0.9895/0.06** |
| 180 | 31.69/0.8820/0.70 | 38.37/0.9853/0.19 | 42.23/0.9881/0.07 | 40.95/0.9845/0.09 | 41.44/0.9876/0.08 | **45.08/0.9941/0.03** |

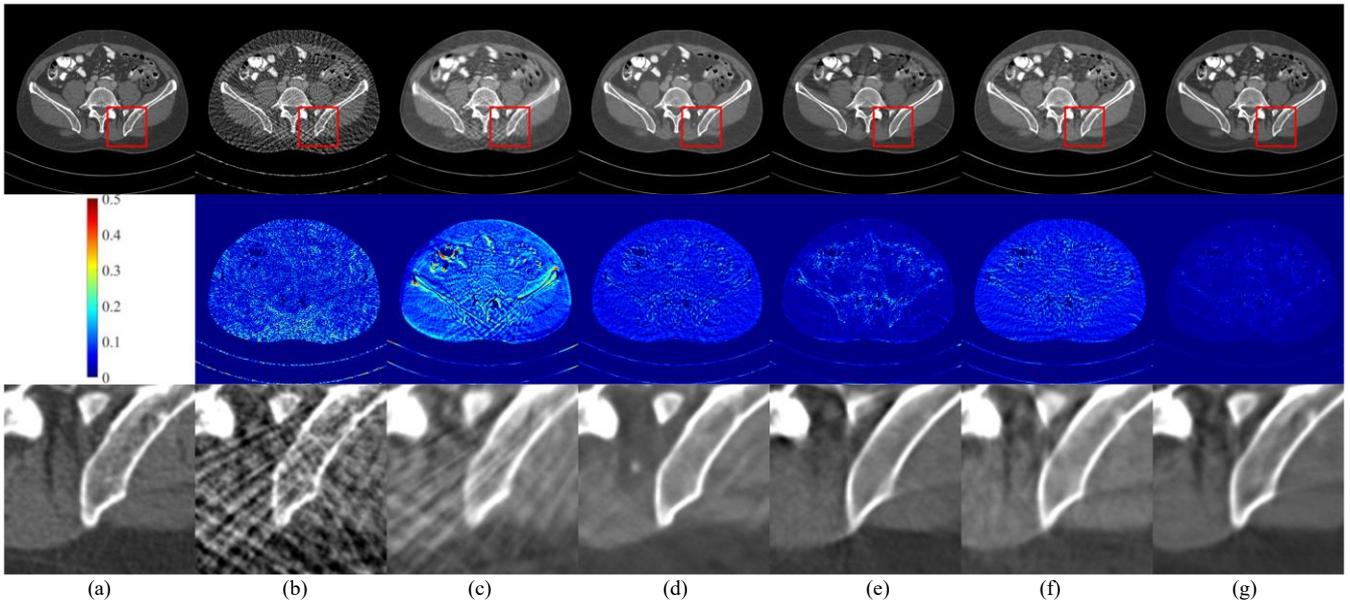

**Fig. 7.** Reconstruction images from 90 views using different methods. (a) The reference image versus the images reconstructed by (b) FBP, (c) U-Net, (d) FBPConvNet, (e) Patch-based DDPM , (f) GMSD, and (g) SWORD . The display windows are [-250,600]. The second row is the difference images.

Moreover, it is evident that the visual effects of SWORD align with the quantitative results. In Figs. 6-7, as the projection view increase from 60 to 90 views, the efficiency of image reconstruction significantly improved. The FBP method produces inferior image quality, with crucial details and structures being indistinguishable (Fig. 6(b)). While the U-Net approach eliminates some artifacts, it results in a loss of important details (Fig. 6(c)). Besides, FBPConvNet is not ideal at recover accurate details (Fig. 6(d)). On the other hand, Patch-based DDPM achieves the second-best noise suppression and structural detail preservation (Fig. 6(e)). However, the results obtained through GMSD exhibit blurred edges (Fig. 6(f)). Remarkably, SWORD produces visual results with the most texture details and the least amount of noise (Fig. 6(g)). Similarly, as shown in Fig. 7, SWORD outperforms other methods, providing more structural information and details. In summary, the results demonstrate that the SWORD method not only excels quantitatively but also yields visually superior results with enhanced texture details and reduced noise compared to the baseline methods.

***CIRS Phantom Reconstruction Results:*** To further validate the robustness of our proposed unsupervised learning scheme, we leverage the prior knowledge learned from the AAPM Challenge Data to evaluate its performance on the CIRS Phantom Data. The results obtained on the CIRS Phantom Data are presented in Table II. Impressively, our method achieves the highest quantitative indices when compared with other comparison methods. It is worth noting that FBPConvNet exhibits notably poor performance on the CIRS Phantom Data. This observation can be attributed to the fact that FBPConvNet is a supervised method, often requiring a substantial amount of data pairs as training sets to learn adequate prior knowledge. On the other hand, Patch-based DDPM and GMSD demonstrate relatively stable results, with slight fluctuations in the quantitative indices. Conversely, Patch-based DDPM and GMSD being the unsupervised methods can adaptively transfer its trained model to various datasets for usage without the need for additional prior knowledge. Our SWORD outperforms Patch-based DDPM and GMSD with higher performance.

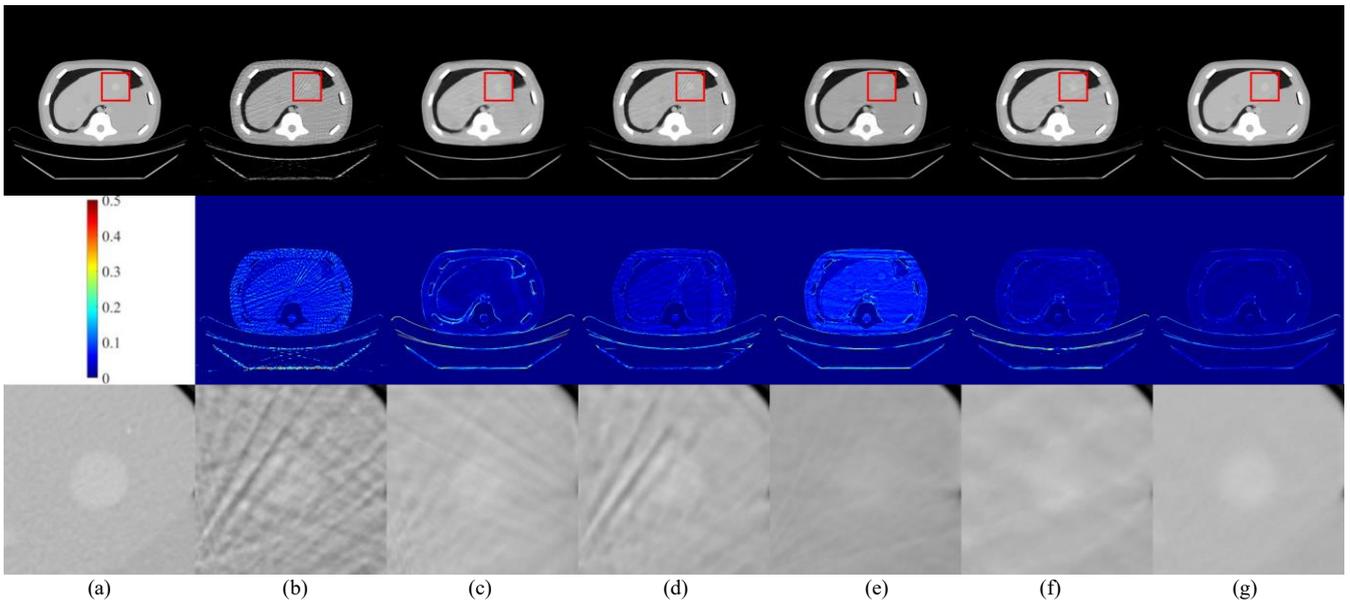

**Fig. 8.** Reconstruction images from 90 views using different methods. (a) The reference image versus the images reconstructed by (b) FBP, (c) U-Net, (d) FBPConvNet, (e) Patch-based DDPM , (f) GMSD, and (g) SWORD . The display windows are [-400,150]. The second row is the difference images.

TABLE II RECONSTRUCTION PSNR/SSIM/MSE(10⁻³) OF CIRS PHANTOM DATA USING DIFFERENT METHODS AT 60, 90, 120, AND 180 VIEWS.

| Views | FBP | U-Net | FBPConvNet | Patch-based DDPM | GMSD | SWORD |
|---|---|---|---|---|---|---|
| 60 | 17.96/0.5028/16.00 | 25.76/0.8883/2.66 | 26.70/0.9333/2.28 | 25.93/0.8979/2.57 | 28.19/0.9209/1.53 | **38.49/0.9778/0.15** |
| 90 | 23.71/0.5982/4.26 | 31.62/0.9542/0.67 | 32.11/0.9579/0.61 | 31.51/0.9676/0.73 | 32.86/0.9603/0.52 | **41.27/0.9862/0.08** |
| 120 | 25.42/0.6827/2.87 | 35.57/0.9727/0.28 | 35.30/0.9730/0.35 | 34.77/0.9781/0.34 | 38.00/0.9815/0.16 | **42.49/0.9895/0.06** |
| 180 | 28.83/0.7994/1.31 | 38.31/0.9852/0.15 | 38.80/0.9848/0.11 | 39.12/0.9895/0.12 | 42.86/0.9911/0.05 | **45.08/0.9941/0.03** |

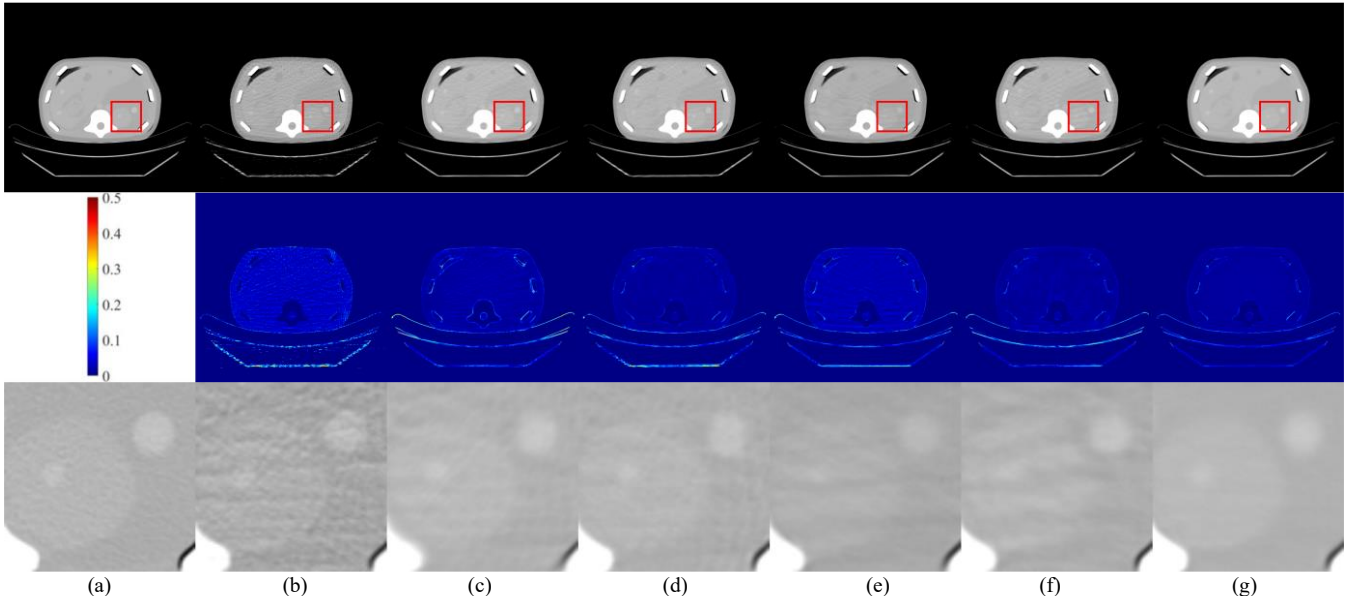

**Fig. 9.** Reconstruction images from 120 views using different methods. (a) The reference image versus the images reconstructed by (b) FBP, (c) U-Net, (d) FBPConvNet, (e) Patch-based DDPM , (f) GMSD, and (g) SWORD . The display windows are [-450,250]. The second row is the difference images.

Simultaneously, the visual effects of SWORD are also exceptional. In Figs. 8-9, we display zoomed regions-of-interest marked by red rectangles. U-Net demonstrates some ability to suppress artifacts but loses image details. FBPConvNet, which combines FBP and deep learning, exhibits significant improvement over FBP in reducing artifacts, but subtle artifacts and blurred internal structures still persist. In comparison to FBPConvNet, the results of the Patch-based DDPM show a slight inferiority. Furthermore, GMSD effectively reduces noise but also sacrifices part of the texture information. In contrast, our SWORD approach demonstrates excellent reconstruction capability and effectively compensates for the artifacts in sparse-view CT. When compared to other competitive reconstruction methods, SWORD excels in terms of noise-artifact reduction and detail preservation, offering the best visual results.

### D. Ablation Study

In study, the proposed SWORD method combines two diffusion models, where the WFDM and WHDM models focus on the full-frequency and high-frequency components. To evaluate the functions of WFDM and WHDM in determining the reconstruction performance, we conduct an ablation study. We examine the quantitative results of WFDM and WHDM, and present their corresponding quantitative results in Table III. It clearly demonstrates the significant improvement in image reconstruction performance achieved by SWORD. The combination of WFDM and WHDM in a serial manner leads to a substantial enhancement in reconstruction quality while preserving their individual strengths.

To further illustrate the impact of the model combination, we provide qualitative results in Fig. 10. It is evident that the sequential combination of WFDM and WHDM significantly promotes the reconstruction quality, showcasing SWORD's ability to effectively leverage the strengths of both models.

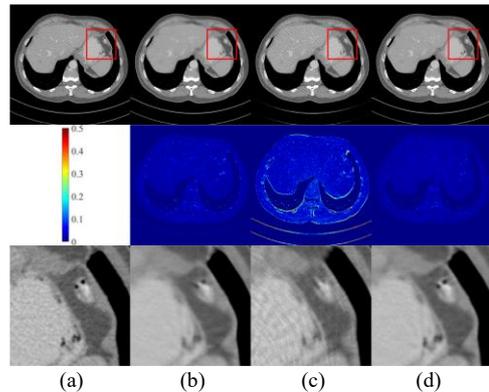

Fig. 10. Reconstructed images obtained from 120 different views employing various methods: (a) Reference image, (b) SWORD without WHDM module, (c) SWORD without WFDM module, and (d) SWORD with full implementation. The display window spans from -450 to 250 Hounsfield units. The second row illustrates the corresponding residuals.

TABLE III RECONSTRUCTION RESULTS OF AAPM CHALLENGE DATA

| Methods | Views | PSNR | SSIM | MSE |
|---|---|---|---|---|
| WFDM | 60 | 37.05 | 0.9700 | 0.23 |
|  | 90 | 39.80 | 0.9814 | 0.11 |
|  | 120 | 41.14 | 0.9862 | 0.08 |
|  | 180 | 43.24 | 0.9907 | 0.05 |
| WHDM | 60 | 24.94 | 0.8008 | 3.26 |
|  | 90 | 27.38 | 0.8668 | 2.24 |
|  | 120 | 29.60 | 0.9077 | 1.13 |
|  | 180 | 34.06 | 0.9588 | 0.40 |
| SWORD | 60 | **38.49** | **0.9778** | **0.15** |
|  | 90 | **41.27** | **0.9862** | **0.08** |
|  | 120 | **42.49** | **0.9895** | **0.06** |
|  | 180 | **45.08** | **0.9941** | **0.03** |

## V. CONCLUSIONS

In this paper, we proposed a novel stage-by-stage wavelet optimization refinement diffusion strategy for sparse-view CT reconstruction, incorporating two diffusion models in the wavelet domain. The key innovation involved the sequential combination of a full-frequency diffusion model and a high-frequency diffusion model. Additionally, we transformed the entire projection dataset into the wavelet domain, allowing us to capture ample prior knowledge. SWORD does not require additional training when applied to different test sets, demonstrating excellent generalization capabilities across diverse datasets. We validated the effectiveness and robustness of SWORD using the AAPM Challenge Data and the CIRS Phantom Data. SWORD excelled in reducing artifacts while preserving structural details and textural perception. Our experimental results clearly demonstrated that the proposed method outperformed the baseline approaches significantly.

In summary, our work introduces a powerful and versatile technique for sparse-view CT reconstruction, showcasing the remarkable capabilities of SWORD in terms of anti-artifact performance and detail preservation.